# Three-Octave Supercontinuum Generation Spanning from Ultraviolet in Lithium Tantalate Waveguides

**LINGFANG WANG,**[1, †, *] **TIANYOU TANG,**[1, †] **HUIZONG ZHU,**[2] **JUANJUAN LU,**[2, *]

[1]School of Optoelectronic Science and Engineering, University of Electronic Science and Technology of China, Chengdu 611731, Sichuan, China
[2]School of Information Science and Technology, ShanghaiTech University, Shanghai 201210, China
[*]Corresponding author: lf.wang@uestc.edu.cn, lujj2@shanghaitech.edu.cn
[†]The authors contributed equally to this work.

**Abstract:** We demonstrate, for the first time, supercontinuum generation spanning more than three octaves in dispersion-engineered thin-film lithium tantalate (TFLT) waveguides. Pumped by a femtosecond laser at 1560 nm, the waveguides yield a spectrum from 240 nm in the ultraviolet to beyond 2400 nm in the near-infrared. The spectral evolution is mapped from low-power harmonic generation (second- and third-harmonic) to a high-power continuum driven by soliton fission and dispersive wave emission. This first demonstration of ultrabroadband nonlinear optics in TFLT establishes it as a competitive, low-loss platform for integrated photonics, with significant potential for applications in frequency metrology and on-chip spectroscopy.

## 1. Introduction

Ultrabroadband supercontinuum generation (SCG) in chip-integrated waveguides has become a cornerstone technology for optical frequency combs [1-2], precision spectroscopy [3-4], and optical clockworks [5-6]. Achieving octave-spanning spectra with low pump thresholds requires materials that combine strong nonlinearities, low optical loss, and tailorable dispersion. To date, thin-film lithium niobate (TFLN) has been a leading platform, leveraging its substantial $\chi(2)$ and $\chi(3)$ nonlinearities within tightly confining waveguides to enable efficient SCG [7-8] and on-chip self-referencing [8-9].

Lithium tantalate ($LiTaO_3$) is an isomorph of $LiNbO_3$, sharing its non-centrosymmetric crystal structure and lower birefringence [10-12]. Crucially, LiTaO3 offers a higher photorefractive damage threshold, lower radio frequency loss tangent and a broader transparency window from ~280 nm to ~5.5 µm [13-14]. Recent advances in thin-film lithium tantalate (TFLT) fabrication have enabled low-loss nanophotonic devices [15-18], yet its potential for generating coherent, ultrabroadband supercontinua remains entirely unexplored.

In this work, we report the first demonstration of supercontinuum generation covering over three octaves in dispersion-engineered, straight TFLT waveguides. By designing the waveguide cross-section to provide anomalous group-velocity dispersion at the 1560 nm pump wavelength, we achieve a spectrum spanning from 240 nm to beyond 2400 nm. We systematically map the spectral evolution across pump powers, revealing a clear transition from a phase-matched $\chi(2)$ harmonic generation regime to a $\chi(3)$-dominated continuum driven by soliton fission and dispersive wave emission. The critical dependencies on input polarization and waveguide geometry are investigated, with experimental results validated by numerical simulations. This work establishes TFLT as a highly promising and versatile platform for integrated ultra-broadband light sources, opening new avenues for on-chip nonlinear photonics.

## 2. Results

The waveguides were fabricated on a commercial 600 nm-thick z-cut TFLT-on-insulator wafer. The fabrication process is outlined in Fig. 1(d). Waveguide patterns were defined using electron-beam lithography and transferred to the TFLT layer via an Ar+-based inductively coupled plasma reactive ion etching process, yielding a sidewall angle of approximately 60° (Fig. 1(b)). Figure 1(c) shows a top-view scanning electron microscope (SEM) image of the final device. A series of straight ridge waveguides with widths ranging from 0.45 µm to 1.65 µm were fabricated to study geometry-dependent effects. No tapers or periodic poling structures were employed.

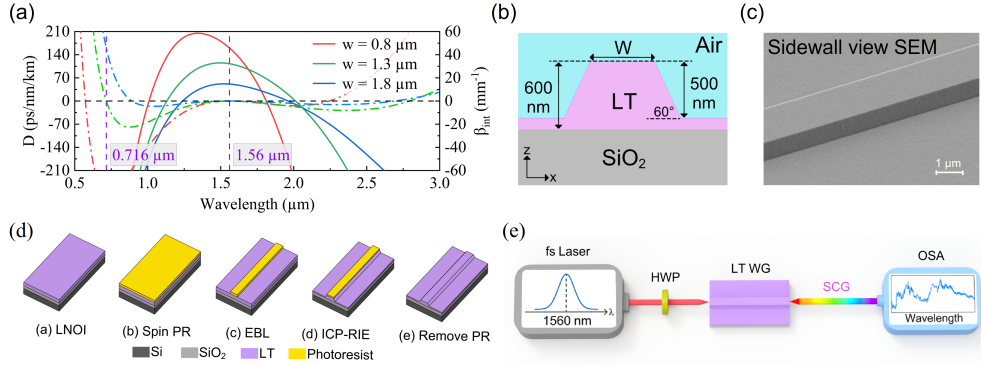

**Fig.1.** TFLT waveguide design, fabrication, and dispersion engineering. (a) Simulated group-velocity dispersion (GVD, solid lines) and integrated dispersion $\beta$int (dashed lines) for the fundamental TE mode in waveguides of widths 0.8, 1.3, and 1.8 µm. The horizontal black dashed line at $\beta$int=0 indicates the phase-matching condition for dispersive wave generation. (b) Schematic cross-section of the waveguide structure. (c) Sidewall-view scanning electron microscope (SEM) image of a fabricated waveguide. (d) Schematic of the fabrication process flow for the TFLT ridge waveguides. (e) Schematic experimental setup.

The group-velocity dispersion (GVD) was simulated using a finite-difference mode solver (COMSOL Multiphysics). Figure 1(a) shows the simulated group-velocity dispersion (GVD, solid lines) for the fundamental TE mode in waveguides with widths of 0.8 µm, 1.3 µm, and 1.8 µm. For widths near 1.3 µm, the waveguide exhibits a region of anomalous GVD at the 1560 nm pump wavelength, which is essential for supporting soliton dynamics.

The phase-matching condition for dispersive wave (DW) generation is governed by the integrated dispersion, $\beta_{\text{int}}(\omega)$ [19], defined as:

$$\beta_{\text{int}}(\omega) = \beta(\omega) - \beta(\omega_0) - (\omega - \omega_0)/v_g(\omega_0) \tag{1}$$

where $\beta(\omega)$ is the propagation constant, $\omega_0$ is the pump frequency, and $v_g$ is the group velocity at the pump frequency. Zero-crossings of $\beta_{\text{int}}$ (dashed lines in Fig. 1(a)) predict the wavelengths for efficient short- and long-wavelength dispersive wave (SWDW, LWDW) emission. The horizontal dashed line at $\beta_{\text{int}}$=0 highlights this condition.

The evolution of the SCG spectrum with on-chip pulse energy was measured in a 1.3 µm wide waveguide under TE polarization. The results are presented in Fig. 2(a). At low pulse energy (~58 pJ), the output spectrum is dominated by distinct, narrowband peaks corresponding to phase-matched second-harmonic generation (SHG) at ~780 nm and third-harmonic generation (THG) at ~520 nm. As the pump power increases, two key phenomena are observed: first, the SHG and THG intensities grow super-linearly; second, a broad spectral shoulder emerges around the SHG wavelength, indicating the onset of dispersive wave (DW) generation via soliton fission in the anomalous dispersion regime.

At a pulse energy of ~231 pJ, the individual SHG peak merges with the broad DW into a continuous, intense band in the visible region, while the spectrum simultaneously broadens dramatically towards both shorter (UV) and longer (infrared) wavelengths. The full spectrum spans from below 240 nm to the detection limit of our spectrometer at 2400 nm, corresponding to a bandwidth exceeding three octaves. This power-dependent evolution clearly illustrates the transition from a perturbative $\chi(2)$-dominated regime to a non-perturbative regime where $\chi(3)$ nonlinearities, soliton dynamics, and dispersive wave emission collectively drive ultrabroadband spectral generation.

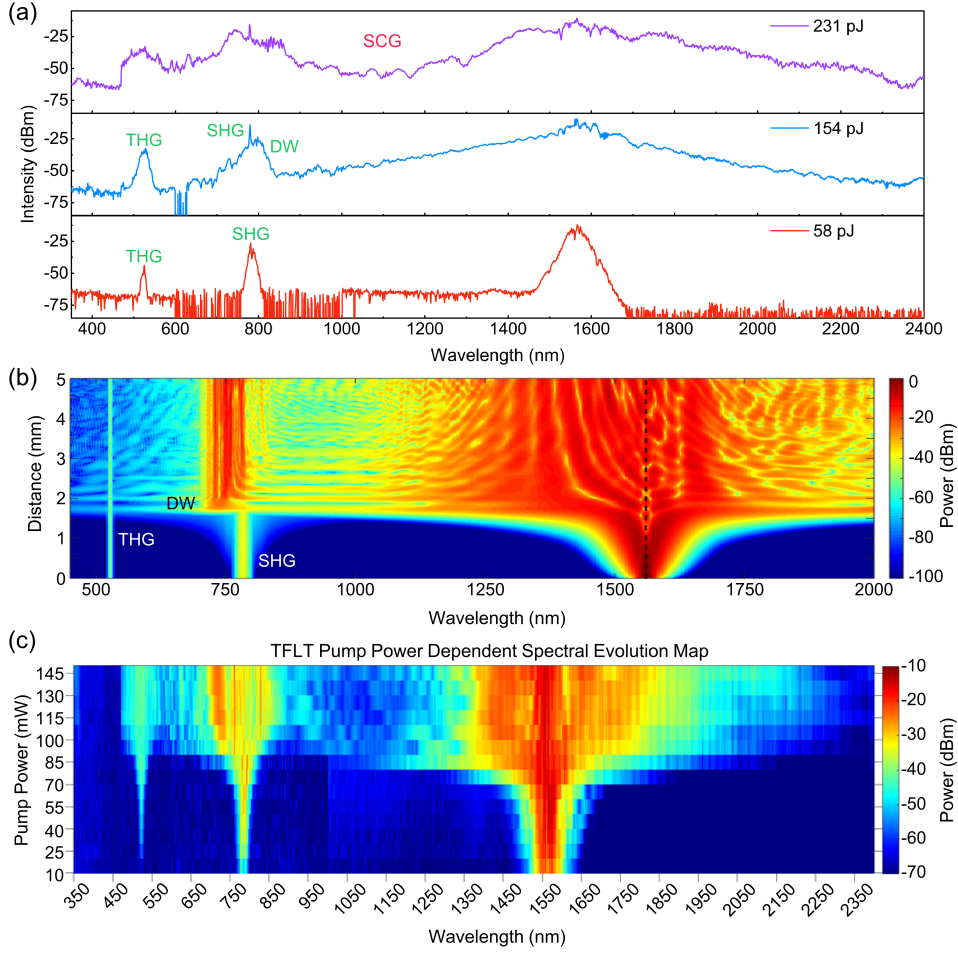

**Fig.2.** Power-dependent experimental spectral evolution and numerical simulation of supercontinuum generation. (a) Measured spectral evolution map as a function of on-chip pulse energy (log scale) in a 1.3 µm wide waveguide under TE polarization. The transition from discrete harmonic generation to a broadband continuum is shown. (b) Corresponding simulated spectral evolution based on the modified GNLSE. (c) Experimental spectral evolution as a function of pump power.

These dynamics were numerically modeled by solving a modified Generalized Nonlinear Schrödinger Equation (GNLSE) that incorporates both second- and third-order nonlinearities, higher-order dispersion, and self-steepening effects:

$$\frac{\partial A(z,t)}{\partial z} + \frac{\alpha}{2} A(z,t) - \sum_{k \geq 2} \frac{i^{k+1}}{k!} \beta_k \frac{\partial^k A(z,t)}{\partial t^k} = i\gamma \left(1 + \frac{i}{\omega_0} \frac{\partial}{\partial t}\right) |A|^2 A + i\kappa A^2 \qquad (2)$$

where, $A(z, t)$ is the pulse envelope, $\beta_k$ are the dispersion coefficients, $\alpha$ is the loss coefficient, $\gamma$ is the effective third-order coefficient, and $\kappa$ models $\chi^{(2)}$ the contribution through an effective nonlinear coefficient dependent on the mode overlap. The simulated spectral evolution (Fig. 2(b)) reproduces the key experimental features, including the initial harmonic peaks and the subsequent DW generation and spectral broadening. Figure 2(c) shows the experimental spectral evolution, illustrating the well agreement with the simulated results.

The full, three-octave-spanning supercontinuum spectrum obtained under on chip pulse energy (~323 pJ) and TE polarization is displayed in Fig. 3(a). The spectrum coherently links the ultraviolet, visible, and near-infrared spectral regions. A pronounced intensity band in the visible is attributed to the combined spectral overlap of the SHG and DW. The near-infrared portion extends smoothly beyond 2400 nm. The high-quality SCG is further confirmed by the ultraviolet spectrum captured down to 240 nm, shown in Fig. 3(b), using a high-sensitivity spectrometer (OPTOSKY ATP5334). The inset photograph in Fig. 3(b) captures the visible light scattered from the waveguide output facet, displaying a vibrant purple-to-red streak—a direct visual manifestation of the generated broadband continuum extends to ultraviolet wavelength. This achievement in a straight, unpoled TFLT waveguide underscores the material's high nonlinear efficiency and the effectiveness of our dispersion engineering.

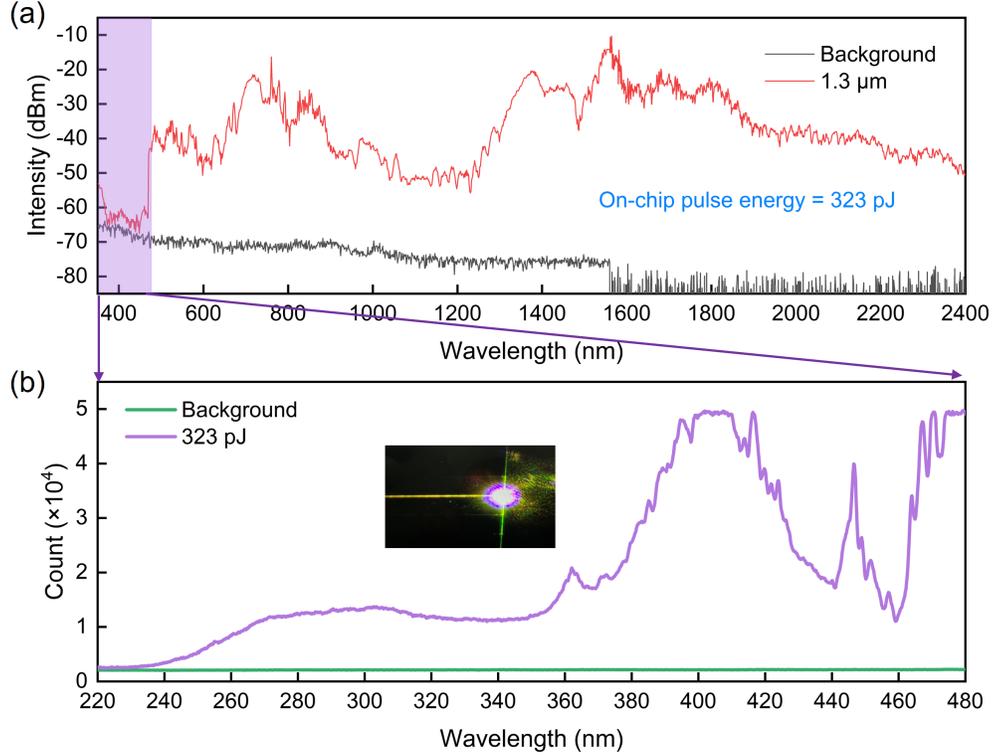

**Fig.3.** Three-octave supercontinuum spectrum and visible output. (a) Full supercontinuum spectrum measured under high pump power and TE polarization, spanning from 240 nm to beyond 2400 nm. (b) Detailed ultraviolet portion of the spectrum (220–480 nm). Inset: photograph of the visible output from the waveguide facet, the purple spot showing the ultraviolet generation.

The SCG process exhibits strong sensitivity to both input polarization and waveguide geometry. To quantify this, we systematically varied the input polarization angle ($\theta$) from 0° (TE) to 90° (TM) at a constant high pump power. The resulting spectral map (Fig. 4(a)) reveals that efficient, broadband SCG with strong DW and harmonic features is sustained only under TE-dominated excitation ($\theta<30°$). Under ~30° to ~60° polarization excitation, the spectral bandwidth collapses, the DW vanishes, and harmonic generation is suppressed. Under polarization from ~60° to ~90°, the resulting spectral are quite similar with the TE-dominated excitation ($\theta<30°$). This behavior can be attributed to the polarization-dependent mode confinement and the different effective $\chi(2)$ tensor elements ($d_{31}$ for TE-like vs. $d_{33}$ for TM-like) accessible in the z-cut crystal orientation. It is worth noting that in lithium tantalate, the ordinary and extraordinary refractive indices ($n_o$ and $n_e$) are relatively close in the telecom band (e.g., $n_o$ ~2.119 and $n_e$~2.123 around 1560 nm), leading to similar dispersion and nonlinear phase-matching characteristics for TE and TM modes. This explains the observed similarity in

SCG bandwidth and nonlinear feature strength between the two polarizations, despite their distinct nonlinear coefficients.

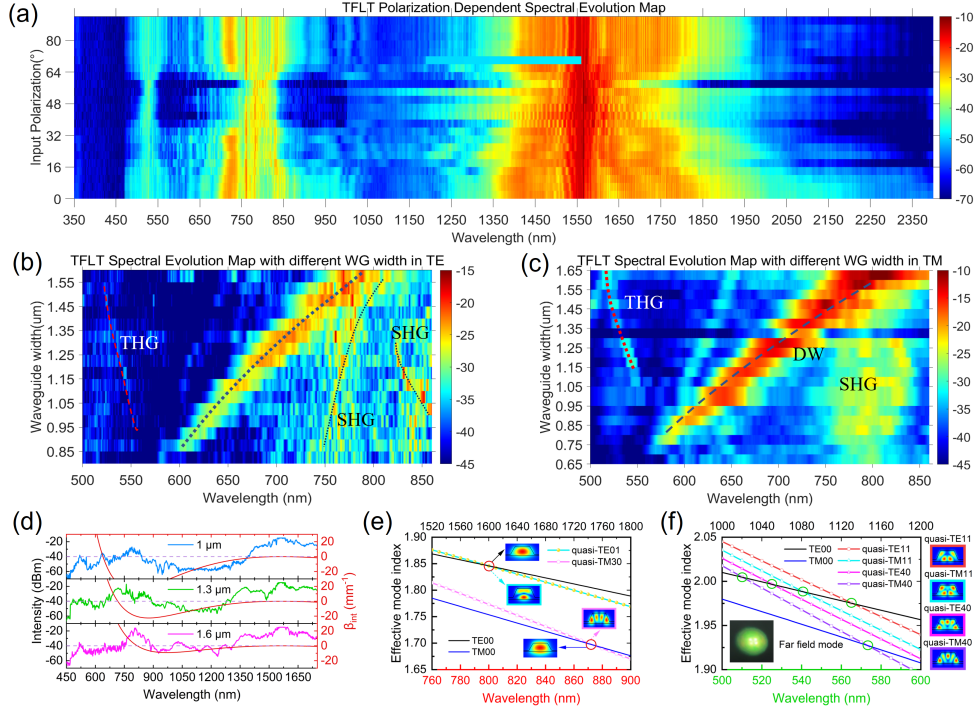

**Fig.4.** Systematic study of polarization/geometry dependence and simulation validation. (a) Spectral map showing SCG dependence on the input polarization angle $\theta$ (0°: TE-like, 90°: TM-like) in a 1.3 μm wide waveguide. (b) Experimental spectra from waveguides of different widths (0.8 μm and 1.6 μm) under TE polarization. (c) Experimental spectra from waveguides of different widths (0.65 μm and 1.65 μm) under TM polarization. (d) Comparison between the measured dispersive wave center wavelength and the simulated phase-matching condition (red line) as a function of waveguide width. (e) Simulated phase-matching wavelengths for SHG versus waveguide width. (f) Simulated phase-matching wavelengths for THG and the corresponding mode profile (inset: far-field mode image experimentally observed near 540 nm, identified as the $TM_{11}$ mode).

The phase-matching conditions for THG, DW, and SHG are intrinsically linked to the waveguide dispersion and thus vary significantly with geometry. Figure 4(b) and (c) show measured spectra from waveguides with different widths under TE and TM polarization, respectively. The central wavelengths and relative intensities of all nonlinear features shift notably with width, demonstrating precise spectral tunability via simple dimensional control. Under TE polarization, broader waveguides (e.g., 1.6 μm) support efficient DW and SHG, while narrower waveguides favor different phase-matching conditions.

To validate our design, we compared experimental results with phase-matching simulations. Figure 4(d) overlays the measured central wavelength of the DW feature with the simulated phase-matching curve derived from $\beta_{int}$ (red lines). The excellent agreement confirms that the observed broad visible feature is a phase-matched DW whose wavelength can be predictably tuned by changing the waveguide width. Figure 4(e) presents the simulated phase-matching wavelengths for SHG (red circles) against waveguide width, and the inset shows the corresponding phase-matched $TE_{01}$ mode profile. The experimentally observed SHG wavelengths under TE and TM excitation (can be find in Figure 4 (b) and (c)) align well with the simulations, verifying their phase-matched origin and the different mode interactions involved. Furthermore, the phase-matching condition for the observed third-harmonic generation (THG) is also investigated. As shown in Fig. 4(f), the simulated THG phase-matching wavelength agrees well with the measured spectral peak around 540 nm. Notably, the far-field mode image captured at this wavelength (inset of Fig. 4(f)) exhibits a clear two-lobe pattern, which matches the simulated profile of the $TM_{11}$ mode, providing direct visual evidence of the phase-matched THG process. This comprehensive agreement between experiment and theory, spanning DW, SHG, and THG, underscores the predictability and control achievable in the TFLT platform.

## 3. Discussion

This work represents, to the best of our knowledge, the first demonstration of supercontinuum generation in thin-film lithium tantalate waveguides. The achieved spectral bandwidth of over three octaves (240 nm to >2400 nm) is exceptionally broad for an integrated nonlinear waveguide. To contextualize our results, Table 1 compares key performance metrics of SCG across several leading integrated photonic platforms.

Table 1 Reported supercontinuum generation across different integrated platforms

| Platform | Pump (nm) | Waveguide Length (mm) | Span Octaves | Short-wave cutoff wavelength | Ref. |
|---|---|---|---|---|---|
| $Si_3N_4$ | 1200 | 300 | ~2 | 400 nm | [20] |
| AlN | 1560 | 8 | ~3.3 | 350 nm | [21] |
| x-cut LN | 950 | 14 | ~1.95 | 350 nm | [22] |
| z-cut LN | 1550 | 6.5 | ~2.7 | 330 nm | [23] |
| GaN | 1560 | 5 | ~2.78 | 570 nm | [24] |
| GaP | 1554 | 3 | ~1.08 | 870 nm | [25] |
| **This work: $LiTaO_3$** | **1560** | **5** | **>3** | **240 nm** | **First SCG in LT** |

Only the demonstrations with relatively smaller short-wave cutoff wavelength of SCG are listed for each different integrated photonic platform.

Our TFLT waveguides achieve a broader bandwidth than previous demonstrations in TFLN and other common platforms, while maintaining a comparable pump wavelength and using a simple, straight waveguide geometry without poling. This superior performance can be attributed to the advantageous material properties of $LiTaO_3$—namely, its high nonlinear coefficients, low optical loss in the measured spectral range, and the precise dispersion engineering enabled by nanoscale fabrication. The ability to generate strong $\chi(2)$ harmonics (SHG, THG) concurrently with a $\chi(3)$-driven continuum is a distinctive feature of this non-centrosymmetric platform, which is absent in centrosymmetric materials like SiN or Si. This synergy, particularly the observed spectral overlap between the dispersive wave and the second harmonic around 780 nm, is promising for applications like *f*-2*f* self-referencing for optical frequency combs.

## 4. Conclusion

We have demonstrated the generation of a supercontinuum spanning more than three octaves in dispersion-engineered nanoscale lithium tantalate waveguides. By pumping in the anomalous dispersion regime at 1560 nm, we achieved spectral coverage from 240 nm to beyond 2400 nm. The nonlinear dynamics were systematically studied, showing a clear evolution from low-power harmonic generation to a high-power continuum mediated by soliton fission and dispersive wave emission. The process is highly efficient under TE polarization and can be precisely tuned via the waveguide width. As the first exploration of ultrabroadband nonlinear optics in TFLT, this work establishes it as a powerful and competitive platform for integrated photonics. The combination of wide transparency, high nonlinearity, low loss, and robust fabrication paves the way for advanced on-chip devices for frequency metrology, broadband sensing, and integrated quantum light sources.

**Funding.** National Natural Science Foundation of China (62305214, 92576113); National Key Research and Development Program of China (2024YFB2807400).

**Acknowledgment.** The facilities used for device fabrication were supported by ShanghaiTech Material Device Lab (SMDL).

**Disclosures.** The authors declare no conflicts of interest.

**Data availability.** The data that support the findings of this study are available from the corresponding author upon reasonable request.